\documentclass[twocolumn, prl, reprint, nofootinbib, showpacs,preprintnumbers,amsmath,amssymb]{revtex4}
\usepackage{varioref,exscale,latexsym,amsmath,amssymb}
\usepackage{graphicx}

\usepackage{graphicx}
\usepackage{slashed}
\usepackage{dcolumn}
\usepackage{bm}
\usepackage{hyperref}
\usepackage{color}
\usepackage{xcolor}

\newcommand{\beq}{\begin{equation}}
\newcommand{\eeq}{\end{equation}}
\newcommand{\bea}{\begin{eqnarray}}
\newcommand{\eea}{\end{eqnarray}}
\newcommand{\nn}{\nonumber}

\def\lsi{\raise0.3ex\hbox{$<$\kern-0.75em\raise-1.1ex\hbox{$\sim$}}}
\def\gsi{\raise0.3ex\hbox{$>$\kern-0.75em\raise-1.1ex\hbox{$\sim$}}}

\def\beq{\begin{equation}}

\def\eeq{\end{equation}}

\begin{document}
\preprint{ACFI-T19-09}

\title{{\bf The arrow of causality and quantum gravity}}

\medskip\

\medskip\

\author{John F. Donoghue${}^{1}$}
\email{donoghue@physics.umass.edu}
\author{Gabriel Menezes${}^{2}$}
\email{gabrielmenezes@ufrrj.br}
\affiliation{
${}^1$Department of Physics,
University of Massachusetts,
Amherst, MA  01003, USA\\
${}^2$Departamento de F\'{i}sica, Universidade Federal Rural do Rio de Janeiro, 23897-000, Serop\'{e}dica, RJ, Brazil}

\begin{abstract}
Causality in quantum field theory is defined by the vanishing of field commutators for space-like separations. However, this does not imply a direction for causal effects. Hidden in our conventions for quantization is a connection to the definition of an arrow of causality, i.e. what is the past and what is the future. If we mix quantization conventions within the same theory, we get a violation of microcausality. In such a theory with mixed conventions the dominant definition of the arrow of causality is determined by the stable states. In some quantum gravity theories, such as quadratic gravity and possibly asymptotic safety, such a mixed causality condition occurs. We discuss some of the implications.
\end{abstract}
\maketitle

When caught making a sign mistake in a phase, a colleague would say: ``Physics does not depend on whether we use $+i$ or $-i$'', and happily change the sign. At first sight, this phrase seems true. Classical fields are real. The probabilities of quantum mechanics are absolute values squared. Measurements in physics do not seem to care if we defined $\sqrt{-1}$ as $+i$ or $-i$.

On second thought, the sign in front of $i$ often does make a major difference. We define time development by
\beq\label{Hamiltonian}
H \psi = i \hbar \frac{\partial}{\partial t}\psi \ \  .
\eeq
This results in ``positive energy'' being defined via $e^{-iEt/\hbar}$.
Canonical quantization is defined via
\beq\label{commutators}
[x, p] = i \hbar ~~~~~~~{\rm or} ~~~~~~[\phi(t,x), \pi(t, x')] = i \hbar \delta^3(x-x').
\eeq
The path integral treatment of quantum physics is defined using $e^{iS}$, not $e^{-iS}$
(in units of $\hbar = c = 1$), with $S$ being the action. The Feynman propagator has very important sign conventions in both the numerator and the denominator, with
\beq\label{propagator}
i D_F(q) =\frac{i}{q^2-m^2 +i\epsilon}
\eeq
with $\epsilon$ infinitesimal and positive. We see that the specific signs in front of $i$ are important in the formalism of quantum mechanics.

On third thought, we can see that these signs are a convention, although they do tie in with another feature of our physical description - in particular the time direction (arrow) of causality.

\begin{figure}[htb]
\begin{center}
\includegraphics[height=28mm,width=85mm]{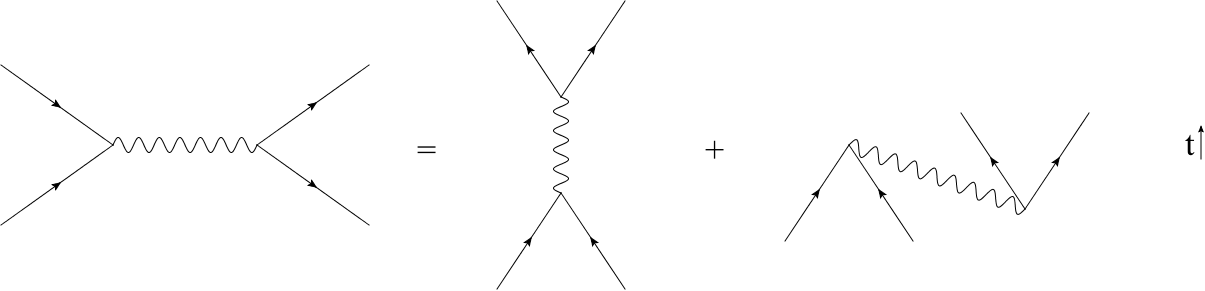}
\caption{The simple Feynman diagram on the left is decomposed into two time ordered diagrams. In one of the time orderings the final particles emerge before the initial particles have annihilated. }
\label{timeorderings}
\end{center}
\end{figure}

In somewhat colloquial language, we would describe causality as ``there is no effect before the cause". In relativistic quantum theories of course, one must be careful about what one means by ``before''. As a simple example, consider the Feynman diagram shown in Figure 1.
The Feynman propagator in coordinate space
includes both forward and backward propagation in time
\beq
iD_{F}(x) = D_{F}^{\textrm{for}}(x) \theta (t) + D_{F}^{\textrm{back}}(x) \theta (-t)
\eeq
with
\beq
D_{F}^{\textrm{for}}(x) = \int \frac{ d^3q}{(2\pi)^3 2E_q} e^{-i(E_qt - \vec{q}\cdot\vec{x})}
\eeq
with $E_q = \sqrt{\vec{q}^2 + m^2}$ and $D_{F}^{\textrm{back}}(x) = (D_{F}^{\textrm{for}}(x))^*$. What we commonly refer to as positive frequency, or $e^{-iEt}$, is propagated forward in time and negative frequency backwards in time. However, we see that in one time ordering the final state particles (the effect) emerge before the initial particles have interacted (the cause). So our colloquial notion of causality is inadequate.

We note in passing that the time advance of the final state vertex is not observable - due to the uncertainty principle. The time difference of the vertices is of order $\Delta t \sim 1/ E$, where $E$ is the center of mass energy. The time localization of the initial state is uncertain by this amount, as is the time resolution of final detection, due to the uncertainty principle.

If you follow the physics back to deeper levels in quantum field theory, you will find that the standard rigorous definition of causality is that operators commute for spacelike separations \cite{GellMann:1954db},
\beq\label{causality}
[{\cal O}(x), {\cal O}(x')] = 0 ~~~{\rm for} ~~(x-x')^2 <0.
\eeq
This restricts the effect of any interactions to processes which occur within the backwards lightcone, i.e. causal processes. However, even here there is an ambiguity. The fundamental definition, Eq. \ref{causality}, does not differentiate between the forward lightcone and the backwards lightcone. By itself it would be equally compatible with causality running in the reverse direction. There is an implied {\em arrow of causality} built into our description of causal processes. It is connected to the signs in front of the various manifestations of $i$ listed above.

We will see that reversing the sign of the factors of $i$ leads to a theory which is causal, but with an arrow of causality which runs from large times to small or negative times.

However, much more interesting is what happens if one allows mixed conventions. It is surprisingly easy for this to occur. For example, if one has higher-derivative kinetic energy terms in the Lagrangian, one obtains a propagator such as
\beq\label{quarticpropagator}
 iD(q) = \frac{i}{q^2- q^4/M^2}= \frac{i}{q^2} - \frac{i}{q^2-M^2}.
\eeq
It turns out that the sign of the propagator is one of the conventions that goes with the ``opposite'' sign conventions (along with the signs of $i\epsilon$ which we will make clear later). Momentum modes near $q^2\sim M^2$ propagate backwards in time compared to the usual modes. We will call these Merlin modes\footnote{The literary reference here is to the wizard in the tales of King Arthur. In some version of the tales, most notably T. H. White's {\it Once and Future King}, Merlin ages backwards in time and derives some of his prophetic power from this fact.  As Merlin says: {\it ``Now ordinary people are born forwards in Time, if you understand what I mean, and nearly everything in the world goes forward too. ($\ldots$) But I unfortunately was born at the wrong end of time, and I have to live backwards from in front, while surrounded by a lot of people living forwards from behind.''} \cite{THWhite}}.
If both conventions are possible and both conventions are present in the same theory, it is then natural to ask what determines the dominant convention. Which convention ``wins'' and provides the macroscopic arrow of causality? Somewhat surprisingly, this turns out to depend on the masses of the particles. In the presence of interactions, the heavy particles decay into light particles. The heavy Merlin particles then do not exist as part of the asymptotic spectrum. They are not part of the asymptotic Hilbert space, which only contains states comprised of the stable particles. The stable particles determine the arrow of causality. However, it is known from the work of Lee and Wick \cite{Lee:1969fy} and Coleman \cite{Coleman} that in theories with quartic propagators our usual ideas of causality are upset, and microcausality violations - of order the inverse width of the heavy particles - occur. An important modern exploration of this phenomenon is given by Grinstein, O'Connell and Wise \cite{Grinstein:2008bg}.

Consider what happens if we define our theory using a path integral with $e^{-iS}$ instead of $e^{iS}$. The generating functional for a scalar field in the presence of a source $J(x)$ for both cases is
\bea
Z_\pm[J] &=& \int [d\phi ] e^{\pm i S(\phi, J)}
\nn\\
&=& \int [d\phi ] e^{\pm i \int d^4 x [\frac12 (\partial_\mu \phi \partial^\mu \phi - m^2 \phi^2 )+ J\phi ]}  \ \ .
\eea
There are various ways that one can add the appropriate $i\epsilon$ to the theory, which are in the end mutually compatible. In the present context, it is easiest and most important to add it to make this Minkowski space Gaussian path integral better defined. To do this, one can add a term
$ \pm  i \epsilon\phi^2/2$ to the Lagrangian density. When combined with the overall factor of $\pm i$, this provides a damping factor $\sim e^{-\epsilon \int d^4 x \phi^2/2}$ in the path integral. If we solve this by the usual ``completing the square'' method,
one obtains
\beq
\hspace{-0.5mm}
Z_\pm [J] = Z[0] \exp\left\{-\frac12\int d^4x d^4 y J(x) ~iD_{\pm F}(x-y) J(y) \right\}
\eeq
with
\beq\label{plusminus}
i D_{\pm F}(x-y) = \int \frac{ d^4q}{(2\pi)^4} e^{-iq\cdot(x-y)} \frac{\pm i}{q^2 - m^2 \pm i \epsilon}   \ \ .
\eeq
The propagator with the plus sign $D_{+F}$ is just the usual Feynman propagator. The other propagator $D_{-F}$ is similar but with different analyticity properties, having poles in the complex plane being across the real axis from the usual case. Using these poles, it can be decomposed into time-ordered components,
\beq
i D_{-F}(x) = D_{-F}^{\textrm{for}}(x) \theta (t) + D_{-F}^{\textrm{back}}(x) \theta (-t)
\eeq
with
\beq
D_{-F}^{\textrm{for}}(x) = \int \frac{ d^3q}{(2\pi)^3 2E_q} e^{+i(E_qt - \vec{q}\cdot\vec{x})}
\eeq
and $D_{-F}^{\textrm{back}}(x) = (D_{-F}^{\textrm{for}}(x))^*$. In contrast with the usual propagator, we see that this form propagates negative frequencies forward in time, and positive frequencies backwards in time. It is the time-reversed version of the usual propagator.

Using the generating functional $Z_- [J]$ and adding interactions, one generates Feynman diagrams for amplitudes, but with all propagators time-reversed from usual. The evaluation of these diagrams proceeds analogously, but with a different Wick rotation to Euclidean momentum space when evaluation the integrals. The usual properties of field theory apply, but with a different arrow of causality. The usual causal behavior is in the direction of an increasing time coordinate. With the signs reversed the causal flow is from positive time to the direction of negative time. This difference is a convention on how we describe the measurement of time\footnote{For example, if you use an hourglass to measure time, the upper chamber measures time by a decreasing amount of sand, and the lower chamber with and increasing amount of sand. We could have adopted the upper chamber definition of the flow of time. Likewise we count down for a rocket launch, and count the years BCE in decreasing order. If we used the convention of temporal flow in the direction of decreasing time, we would be using the $-$ convention for propagators.}.

This difference makes sense from the point of view of time-reversal symmetry. The time-reversal operation is {\em anti-unitary}, as it involves complex conjugation as well as field transformations. While the Lagrangian may be time-reversal invariant, the full path integral is not because we form it using $e^{iS}$. Under time-reversal, $Z_+$ is turned into $Z_-$, reversing the direction of the arrow of causality. Both versions are equally causal, but with different directions of causality.  We also note that the ``thermodynamic arrrow of time'', which describes the direction of the increase of entropy, follows the direction of causal processes and would reverse under the full time-reversal of the path integral.

Canonical quantization also carries an implicit time direction. This can be seen for example by relabeling the time coordinate via $t=-\tau$. In this case we see that Hamiltionian evolution changes from Eq. \ref{Hamiltonian} to
\beq
H \psi = - i \hbar \frac{\partial}{\partial \tau}\psi \ \ .
\eeq
Likewise, the canonical commutators, Eq. \ref{commutators}, change sign to
\beq
[\phi(t,x), \bar{ \pi}(t, x')] = -i \hbar \delta^3(x-x')~~~~{\rm with} ~~~~\bar{\pi}= \frac{\partial {\cal L}}{\partial (\partial_\tau \phi)}.
\eeq
Despite the unconventional signs, this is the exact same theory, just with a change of the clock direction.
Our standard formalism for quantum theory has been built with the knowledge that we want the direction of causality to be in the direction of increasing time, and this fact is reflected in the various factors of $i$.
{\it There is a causal direction built into our laws of quantum physics}.

If there are particles with both sign conventions in the asymptotic spectrum, there can be no macroscopic notion of causality. However,
if there are unstable particles of a finite lifetime with the opposite sign convention, then causality can hold on macroscopic scales while being violated on short time scales. While this seems like an odd situation to consider, it does happen in a very simple manner if a theory has higher derivatives in the action, such as
\beq\label{scalar}
{\cal L} = \frac12 \left[\partial_\mu \phi \partial^\mu \phi -m^2\phi^2\right] - \frac{1}{2M^2} \Box \phi \Box  \phi.
\eeq
As noted above in Eq. \ref{quarticpropagator}, this is equivalent to adding a particle with a negative kinetic energy.
Note that reversing the sign of the kinetic energy,
\beq \label{ghosts}
Z \sim \int [d\phi] e^{i \int d^4x [-\frac12 \partial_\mu \phi \partial^\mu \phi \, + \, \cdots ] }   ,
\eeq
is equivalent to using $e^{-iS}$, and these particles will carry the opposite arrow of causality.

However, heavy particles are most often unstable and decay into lighter ones. This can be seen in the propagator, after the inclusion of the self-energy quantum correction. For our scalar theory we would have
\beq
i D_F(q) = \frac{i}{q^2 -m^2 + \Sigma(q) - q^4/M^2}
\eeq
where $\Sigma (q) $ is the self energy. Above some threshold $\lambda_T$ for producing light decay products, the self energy develops an imaginary part $\textrm{Im} ~\Sigma(q) = \gamma(q)$ for $q^2 > \lambda_T^2 \geq 0$. Unitarity requires $\gamma(q) \geq 0$. Now take $m^2 \ll M^2$. If the threshold occurs above the mass $m$ (absorbing $\textrm{Re} ~\Sigma $ into $m^2$) then the scalar particle is stable. If the threshold is below $q^2=m^2$ then it is a normal resonance
\beq
iD_F \sim_{q^2\sim m^2} = \frac{i}{q^2-m^2 +i\gamma_m}
\sim \frac{i}{q^2 - \left(m_r - i\frac12 \Gamma\right)^2}
\eeq
where $\gamma_m =\gamma(m^2)$. However, there is a heavier second resonance near $q^2= M^2$, where we find
\begin{eqnarray}  \label{merlin}
iD_F(q) &=& \frac{i}{q^2 - \frac{q^4}{M^2} + i \gamma(q) } \nonumber \\
&=& \frac{i}{\frac{q^2}{M^2}[M^2 - q^2 + i \gamma(q) (M^2/q^2)]} \nonumber  \\ &\sim& \frac{-i}{q^2-M^2 -i \gamma_M} \ \ .
\end{eqnarray}
Comparison with Eq. \ref{plusminus}, shows that this is a finite-width version of the time reversed propagator $iD_{-F}$. This is the more precise definition of Merlin modes. Not only is there a minus sign in the numerator\footnote{This numerator minus sign would qualify it for the more generic name ``ghost''.} but there is also a minus sign in front of the width. We have a separate paper proving that these modes do not upset the unitarity nor stability of the theory \cite{unitarity}. The dominant arrow of causality in such a theory is provided by the stable modes, but there can be reversed time propagation over timescales of order the inverse width of the Merlin modes.

We could create higher derivative versions of any particular quantum field theory. One reason for doing so is that it converts a renormalizable theory into a finite theory. Extra powers of momenta in the denominator damp out high momentum propagaton and remove UV divergences. This was the original motivation of Lee and Wick \cite{Lee:1969fy}. However, the higher-derivative extension is essentially obligatory for quantum gravity. Despite the historical complaints about their union, at low energies quantum field theory and general relativity work well together as a perturbative effective field theory \cite{Donoghue:1994dn}. However, the nature of the effective field theory requires that loop processes generates higher-derivative interactions. The gravitational vertices in relativistic theories are proportional to the energy-squared, and power counting tells us that one loop diagrams carry powers of $E^4$ which is equivalent to fourth-order derivative effects. In particular, loops of ordinary matter fields generate curvature-squared terms in the action\footnote{These are formally divergent and need to be renormalized. However, even if the ultimate theory is finite, these loops would leave residual finite curvature-squared terms in the action.}. The curvatures are second order in derivatives, so that curvature-squared terms are fourth order in derivatives.

Will these higher derivatives create causality violations due to their conflicting arrows of causality? Not necessarily. It could be that the effective field theory falls apart at energies well below those where the Merlin modes would be excited. In this case, the gravitational theory could have new degrees of freedom and new interactions and could perhaps be fully causal. However, there are also approaches to quantum gravity, such as quadratic gravity and asymptotic safety, which do keep the usual gravitational degrees of freedom at all energies and do have interesting causal properties. We describe these below. Moreover, the case can be made that all gravitational theories display some {\em causal uncertainty}, due to their spacetime fluctuations. As an example of this we note that even in the effective theory, quantum corrections to the trajectories of massless particles are spin dependent and the particles no longer follow geodesic motion \cite{Bjerrum-Bohr:2016hpa}. This makes the concept of a lightcone uncertain. We will explore this idea of causal uncertainties in a subsequent paper \cite{uncertainty}.

Quadratic gravity \cite{Stelle:1976gc, Salvio:2018crh, Donoghue:2018izj} is an approach that keeps the curvature squared terms in the fundamental action at all energies. It is a renormalizable \cite{Stelle:1976gc} and unitary \cite{unitarity}  quantum field theory. The fundamental propagators do involve terms quartic in the momenta, and hence will have the Merlin modes. Indeed, the scalar part of the spin-two propagator has the form
\bea\label{spintwo}
D_2(q) &=&  \Biggl[ {q^2+i\epsilon}- \frac{\kappa^2 q^4}{2\xi^2(\mu)}
\nn\\
&-& \frac{\kappa^2 q^4 N_{\textrm{eff}}}{640\pi^2} \left( \ln \left(\frac{|q^2|}{\mu^2}\right)-i\pi \theta(q^2)\right)
\Biggr]^{-1}
\eea
where $1/3\xi^2$ is the coefficient in front of $R^2- 3 R_{\mu\nu} R^{\mu\nu}$ in the action, and $N_{\textrm{eff}}$ is the effective number of light particles. This produces a Merlin resonance near $m_r^2 = 2 \xi^2 /\kappa^2 \sim \xi^2 M_P^2$.

Asymptotic safety \cite{Niedermaier:2006wt} is an approach which defines the gravitational action by using the renormalization group to integrate the quantum corrections in a Euclidean path integral from a fixed point at high energy down through all the scale to zero momentum. When continued to Lorentzian space, it yields an action which contains an infinite series of all possible Lagrangians consistent with general covariance. The infinite number of coefficients are in principle fixed by a few parameters at the UV fixed point. The propagators of the final Lorentzian theory then contain all powers of the momenta. In practice, the theory is explored by truncating the action to a finite basis. In any such truncation, the partial fraction decomposition would reveal the presence of Merlin modes. While it is hard to extrapolate from a truncation of an unknown infinite series, it is then likely that this would also be a property of the full theory.

The phenomenology of theories with mixed causal arrows has been partially explored and in principle there are potentially observable consequences. For example, Grinstein, O'Connell and Wise \cite{Grinstein:2008bg} considered the evolution of a wavepacket in a scalar Lee-Wick theory and showed that there is a signal that arrives in advance of causal expectation. In addition, Ref. \cite{Alvarez:2009af} considered the emergence of final particles at times earlier than the scattering vertex at the LHC, which would occur in a higher-derivative extension of the Standard Model. A related phenomena is the advance of the scattering phase through a resonance in the reverse direction. In potential scattering, Wigner showed that the derivative of the scattering phase is related to the resonance time delay \cite{Wigner}. In normal resonances, this leads to the phase of elastic scattering moving counterclockwise through $90^o$ on the Argand diagram. In quadratic gravity we have calculated the unitary amplitude for scattering in the spin-two channel \cite{Donoghue:2018izj}
\beq\label{t2amplitude}
T_2(s) = - \frac{N_{\textrm{eff}} s}{640 \pi}\,\bar{D}_2(s)
\eeq
where $D_2$ is the propagator of Eq. \ref{spintwo}. This produces a resonance phase moving clockwise through $90^o$.

In these cases, the diagnosis is straightforward from what we have already seen of such theories. Referring back to Fig. 1, we note that the excitation of a normal resonance occurs in the forward time direction as seen in Fig. 1a. After a lifetime proportional to the inverse width, the decay products emerge. However, a Merlin resonance propagates backwards in time as in Fig. 1b, by an amount also proportional to the inverse width. If the backwards lifetime is large enough, one could potentially detect the early emergence of decay products. However, for gravity the timescales are far too short for observation, as they are proportional to the Planck scale, which as a time unit is $t_p\sim 10^{-43}$ sec.

\begin{figure}[htb]
\begin{center}
\includegraphics[height=15mm,width=65mm]{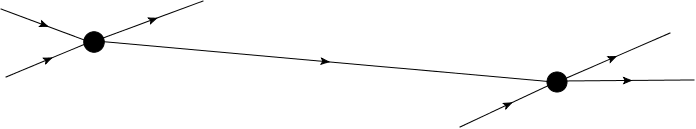}
\caption{The first scattering process on the left contains some causality violation, which limits our ability to be precise about the causal properties of the second scattering process on the right. }
\label{uncertainty}
\end{center}
\end{figure}

Moreover, there is an intrinsic causal uncertainty in these tests also. The ideas of wavepackets or precisely defined beams and detectors are idealized concepts. Forming a wavepacket or producing a beam are themselves done by previous scattering processes. In a theory with mixed causal arrows, these will also have their causal mismatches. An example is illustrated in Fig. 2. Much like the usual uncertainty principle limits our ability to see the early vertex in Fig. 1b, the previous mixed causal processes will limit our ability to produce a sharp wavepacket or beam to test the nature of causality violation.

The arrow of causality is a more precise concept than the arrow of time.  Causality is a specific microscopic phenomenon, a property of the fundamental scattering amplitudes. Discussions of the arrow of time often state that the microscopic laws of physics are the same with time running forwards or backwards. That is true for the basis of classical physics, which follows from the minimization of a Lagrangian. However as discussed above it is not true for quantum physics if you also include the quantization procedure in the phrase ``laws of physics''. If we define our quantum theory by a path integral with $e^{iS}$ this defines the direction of the arrow of causality. To reverse the arrow of causality, we would have to define a different quantum theory using $e^{-iS}$. This is because the time reversal operation is anti-unitary. Reversing the arrow of causality will also reverse the thermodynamic arrow of time because the increase of entropy occurs in the direction that causal processes occur.

Moreover, it is important to recognize that the arrow of causality can potentially be violated. In theories with higher derivatives, there can be modes which briefly propagate against the dominant arrow of causality, which is set by the stable states of the theory.  Gravity represents the most likely situation for this to occur. While the acausal properties of gravitational scattering are beyond reach of observation, it would be interesting to study the effect of the causal uncertainty in the early universe.

\section*{Acknowledgements} We would like to thank J. Fr\"{o}hlich, B. Holdom, B. Holstein, A. Salvio and A. Tolley for comments and conversations. The work of JFD has been partially supported by the US National Science Foundation under grant NSF-PHY18-20675 (JFD). The work of GM has been partially supported by  Conselho Nacional de Desenvolvimento Cient\'ifico e Tecnol\'ogico - CNPq under grant 310291/2018-6 and Funda\c{c}\~ao Carlos Chagas Filho de Amparo \`a Pesquisa do Estado do Rio de Janeiro - FAPERJ under grant E-26/202.725/2018.

 \end{document}